\documentclass{iopart}
\newcommand{\be}{\begin{equation}}
\newcommand{\ee}{\end{equation}}
\usepackage{graphicx}

\begin{document}
\title{Exact Static Solutions for Scalar Fields Coupled
to Gravity in $(3+1)$-Dimensions}
\author{Ayse H. Bilge and Durmus Daghan}

\address{Istanbul Technical University, Faculty of
Science and  Letters, Mathematics Engineering Department, Maslak,
TR-34469 Istanbul, TURKEY} \ead{bilge@itu.edu.tr,
daghand@itu.edu.tr}

\begin{abstract}
Einstein's field equations for a spherically symmetric metric
coupled to a massless scalar field are reduced to a system
effectively of second order in time, in terms of the variables
$\mu=m/r$ and $y=(\alpha/ra)$, where $a$, $\alpha$, $r$ and $m$
are as in [W.M. Choptuik, ``Universality and Scaling in
Gravitational Collapse of Massless Scalar Field", \textit{Physical
Review Letters} {\bf{70}} (1993), 9-12]. Solutions for which $\mu
$ and $y$ are time independent may arise either from scalar fields
with $\phi_t=0$ or with $\phi_s=0$ but $\phi$ linear in $t$,
called respectively the positive and negative branches having the
Schwarzschild solution characterized by $\phi=0  $ and
$\mu_s+\mu=0$ in  common. For the positive branch we obtain an
exact solution which have been in fact obtained first in [I.Z.
Fisher,``Scalar mesostatic field with regard for gravitational
effects", \textit{Zh. Eksp. Teor. Fiz.} {\bf{18}} (1948), 636-640,
gr-qc/9911008] and rediscovered many times (see D. Grumiller,
``Quantum dilaton gravity in two dimensions with matter", PhD
thesis, \textit{Technische Universit$\ddot{a}$t, Wien} (2001),
gr-qc/0105078) and we prove that the trivial solution $\mu=0$ is a
global attractor for the region $\mu_s+\mu>0 $, $\mu<1/2$. For the
negative branch discussed first in [M. Wyman, ``Static spherically
symmetric scalar fields in general relativity", \textit{Physical
Review D} {\bf{24}} (1981), 839-841] perturbatively, we prove that
$\mu=0$ is a saddle point for the linearized system, but the
non-vacuum solution $\mu=1/4$ is a stable focus and a global
attractor for the region $\mu_s+\mu>0$, $\mu<1/2$.
\end{abstract}



\section{Introduction}
The initial value problem for Einstein's equations with massless
scalar field  was studied analytically by Christodoulou in
\cite{Ch86a} and \cite{Ch86b} where it was shown that ``small
initial data" disperses while for  ``large initial data" the end
state is a black hole surrounded by vacuum. The work of  Choptuik
\cite{C93} on the  numerical search for ``critical initial data"
that would separate these two types of behavior  led to the
discovery of the ``threshold phenomenon". For various one
parameter families of initial data, where $p$ denotes  the
amplitude or the width of pulses of various shapes, it was
observed that for $p\ll p_*$ the end state of the evolution is a
flat space-time (external field dissipates), while for $p\gg p_*$
the end state of the evolution is a black hole.  For $p$ near
$p_*$, where the black hole formation first starts, the black hole
mass is proportional to $(p-p_*)^\gamma $ where the constant
$\gamma$ is nearly equal to $0.37$   independently of the initial
data. Furthermore, the critical solution exhibits a universal
scaling behavior in the sense that it is periodical in suitable
time and space coordinates with a period nearly equal to  $3.44$
for all families. Threshold behavior was studied afterwards for  a
number of  configurations and similar types of critical behavior
have been observed. A detailed overview of the literature on the
threshold phenomena can be found for example in  \cite{GN03}.

The field equations used in  numerical studies in the literature
form a consistent but overdetermined system. In this paper we
shall make use of the complete set of field equations as in
\cite{H04} to decouple the equations for the metric functions from
the equations for the scalar field. These equations form a normal
system and they are effectively of second order in time.   In
logarithmic coordinates, they have no explicit dependency on the
independent variables, as expected from the symmetry analysis of
the original equations. The second order hyperbolic equation for
the mass density involves a parameter that can be either plus or
minus one, hence the decoupled system has positive and negative
``branches". Besides its suitability for perturbative studies, in
the time independent case, this system reduces to an autonomous
second order system and allows a rigorous analysis of the negative
branch, which is not transparent in the original formulation.


When the field equations are written in that form, a non-flat,
special solution for the negative branch is obvious: $\mu={1\over
4}$, or $m={1\over 4}r$, for which the  external field  is linear
in time. Thus surprisingly, time varying external fields give rise
to static  metrics.  A perturbative study of the full system is
deferred to a future work.  In Sections 4 and 5 we restrict
ourselves to the static case, which can be realized either as the
ones corresponding to time invariant external fields or the ones
with $\phi_s=0$ but $\phi_t\ne 0$. For the positive branch we
obtain an exact analytic solution while for the negative branch,
we prove that all solutions starting in the region $\mu_s+\mu>0$,
$\mu<1/2$,  stay in that region and the constant solution
$\mu={1\over 4}$ is a global attractor. Although  there is no
non-vacuum static black hole solution, for large values of
$\phi_s$ of $\phi_t$, $\mu$ gets very close to $1/2$ and decays
rapidly to negative values, as depicted in Figure 3b. The trivial
solution $\mu=0$ and the Schwarzschild  solution with
$\mu_s+\mu=0$ are of course common the  both branches.

It was pointed out by D. Grumiller in a private communication that
the solution for the positive branch dates back to Fisher
\cite{F48}, as discussed in detail in \cite{Gr1} and \cite{Gr2},
Fisher's solution have been  rediscovered in
\cite{Berg57}-\cite{Wy81} and errors of the original equations
28-29 in \cite{F48} were corrected by Grumiller \cite{Gr1}. In
addition we may cite the works of Bronnikov and Shikin \cite{Br02}
on the characterizations of the global properties of static,
spherically symmetric configurations in various dimensions. As
opposed to the popularity of the positive branch, the second class
of static solutions that we call the negative branch is first
noticed  by Wyman \cite{Wy81} where it is studied perturbatively.

\section{The Field Equations}

Let $M$ be a four dimensional Lorentzian manifold and $g$ be a
spherically symmetric metric on $M$.  In the coordinate system
adopted in  \cite{C93}, the local coordinates are
$x^\mu=(t,r,\theta,\varphi)$ with
$$0<\theta<\pi,\quad  0<\varphi<2\pi, \quad r>r_o>0, \quad  t>t_o>0.$$
The metric is given by
$$ds^2=-\alpha^2\ dt^2 +a^2 \ dr^2 + r^2\ d\theta^2 + r^2
sin^2\theta  \ d\varphi^2\eqno(2.1)$$ where $\alpha=\alpha(r,t)$,
$a=a(r,t)$.

The Einstein's equations coupled to a scalar, static particle are
\cite{W84}
$$R_{\mu\nu}-\frac{1}{2}g_{\mu\nu}R=\kappa
T_{\mu\nu},$$ where $$T_{\mu\nu}=\partial_{\mu}\phi
\partial_{\nu}\phi- \frac{1}{2}g_{\mu\nu} (\partial_{\lambda}\phi
\partial^{\lambda}\phi)$$
and $\phi$ satisfies the wave equation
$$g^{\mu\nu} \nabla_{\mu} \nabla_{\nu} \phi =
(-g)^{-\frac{1}{2}}\partial_{\mu}[(-g)^{\frac{1}{2}}g^{\mu\nu}
\partial_{\nu}\phi]
$$ with $g=\det( g_{\mu\nu})$ \cite{BD82}.
The complete set of field equations are given below, where we have
written $\kappa=8\pi$ and omitted an equation which was
identically satisfied.
$$
{a_t \over a} = 4\pi r \phi_t\  \phi_r,\eqno(2.2a)$$
$${a_r \over a} - {\alpha_r\over\alpha}+{a^2-1\over r}=0,\eqno(2.2b)$$
$${a_r \over a} +{\alpha_r\over \alpha}=
  4\pi r \left[ \phi_r^2 +{a^2\over
  \alpha^2}\phi_t^2\right],\eqno(2.2c)$$
$$\left({a\over \alpha} \phi_t\right)_t={1\over r^2}
\left(r^2 {\alpha\over a}\phi_r\right)_r.\eqno(2.2d)$$

\vskip 0.3cm \noindent {\bf Remark 2.1.} The system above is
over-determined but integrable in the sense that the equations for
$a_t$ and $a_r$ are compatible, as noted in \cite{H04}. The degree
of freedom in the system appears to be three functions of a single
variable, corresponding to the initial data for $\phi$, $\phi_t$
and $\alpha$. This last freedom is  actually redundant, since it
can be eliminated by redefining $t$. The use of the complete
system will allow the decoupling of the equations for $a$ and
$\alpha$ from the equation for $\phi$. \vskip 0.3cm

The literature on the threshold phenomenon works with  the system
of equations (2.2b-d), starting with initial data for $\phi$.  We
shall use the  function $m(r,t)$ defined in terms of $a(r,t)$ as
our primary variable and introduce logarithmic coordinates to
obtain a normal (i.e. not overdetermined) system with no explicit
dependency on the independent variables.

We start by defining the  variables
$$z={\alpha\over a},\quad\quad a^2=\left[1-{2m\over r}\right]^{-1}.
\eqno(2.3)$$ Note that as $r>0$, we need to impose \ $r-2m>0$\  to
ensure the positivity of $a^2$.  The equations (2.2a-d) can be
rearranged in the form
$${z_r\over z}={2m\over r(r-2m)},\eqno(2.4a)$$
$$ {m_t\over r(r-2m)}=4\pi \phi_t\phi_r, \eqno(2.4b)$$
$${m_r\over r(r-2m)}= 2\pi
\left[ \phi_r^2+{\phi_t^2\over z^2}\right],\eqno(2.4c)$$
$$\left({1\over z} \phi_t\right)_t={1\over
r^2} \left(r^2 z\phi_r\right)_r.\eqno(2.4d)$$ Finally passing to
logarithmic coordinates we obtain an autonomous system, as given
below in Proposition 2.1.

\vskip 0.3cm \noindent \textit{{\bf \textit{Proposition 2.1.}} Let
$$s=\ln(r),\quad z=ry,\quad m=r\mu.\eqno(2.5)$$
Then, the  system of equations (2.2a-d) is equivalent to the
system of equations
$${y_s\over y}=-2+{1\over 1-2\mu},\eqno(2.6a)$$
$$\frac{1}{2\pi} \frac{\mu_t}{1-2\mu}=2
\phi_s \phi_t,\eqno(2.6b)$$
$$\frac{1}{2\pi} \frac{\mu_s+\mu}{1-2\mu}=
 \phi_s^2 +{\phi_t^2\over y^2},\eqno(2.6c)$$
$$\left({\phi_t\over y}\right)_t=2y\phi_s+
(y\phi_s)_s,\eqno(2.6d)$$ where  $\mu_s+\mu $ and $\mu_t/y$ are
constrained by
$${\mu_s+\mu\over 1-2\mu}\ge 0,\quad \quad {1\over 2\pi}
  {(\mu_s+\mu)\pm (\mu_t/y )\over 1-2\mu}=
\left(\phi_s\pm {\phi_t\over y}\right)^2\ge 0.\eqno(2.7)$$} \vskip
0.3cm

\noindent {\bf Remark 2.2.} In addition to the translation
invariance, the system (2.6a-d) admits the symmetry
$$(t, y)\quad \to \quad  \left(\tau=\tau(t),\tilde{y}={1\over
d\tau/dt} y\right).\eqno(2.8)$$ Thus $y$ can be scaled by an
arbitrary function of $t$ only, hence its time evolution is
determined in terms of the time evolution of $\mu$.

\vskip 0.3cm \noindent {\bf Remark 2.3.} A special case of (2.8)
is the transformation
$$(t,y)  \to (\tau=\ln t, \tilde y = ty).\eqno(2.9)$$
Hence one can use $(\tau,\tilde y)$ as well as $(t,y)$ in
(2.6a-d). If $\tilde{y}$ is periodic in $\tau$, i.e.,
$\tilde{y}(\tau,s)=\tilde{y}(\tau+T,s)$. Then
$$e^{\tau}y(e^\tau,s)=e^Te^\tau y(e^Te^\tau,s),$$
or
$$y(t,s)=e^T y(e^Tt,s),$$
hence periodic solutions lead to the ``echoes" of the original
solution. \vskip 0.3cm

The regions determined by the conditions $\phi_s=\pm {\phi_t\over
y}$ and $\mu_s+\mu=\pm {\mu_t\over y}$ are shown in the Figures 1a
and 1b respectively.    As the equations (2.6a-d) are invariant
under the discrete symmetry $\phi_s\rightarrow-\phi_s$ it is
sufficient to work with $\phi$ in regions $I$ and $II$ for
example. Also for $1-2\mu>0$ only region $I^*$ in Figure 1b is
allowed.  As it will be seen in the next section, the half plane
consisting of the regions $I$ and $II$  in the $(\phi_s,\phi_t/y)$
plane is mapped to the quarter plane $I^*$ as a double cover. We
shall call these as the ``positive" and ``negative" branches. The
boundary of the regions $I$ and $II$ is characterized by
$\phi_s=\pm \phi_t/y$  or equivalently with $\mu_s+\mu=\pm
\mu_t/y$. The decoupled equation (3.1b) below admits non-vacuum
solutions $\mu_s+\mu=\pm \mu_t/y$, but when we take into account
the wave equation for $\phi$ we get $\phi_s=\phi_t/y=0$, hence
$\mu_s+\mu=\mu_t=0$.

\vskip 0.3cm
\section{Reduction of the field equations}

We shall eliminate $\phi$ from (2.6a-d) and obtain a system of
equations for $\mu $ and $y$.  The proof of the proposition below
is  essentially a straightforward computation. We have introduced
the auxiliary variables $X$ and  $Y$ below in order to make the
proof more readable.

\vskip 0.3cm \noindent \textit{{\bf \textit{Proposition 3.1.}} For
$\phi_s^2- (\frac{\phi_t}{y})^2\ne 0$ the system of equations
(2.6a-d) is equivalent to the system of equations given below.
$${y_s\over y}=-2 +{1\over (1-2\mu)},\eqno(3.1a)$$
$$\mu_{ss}+\mu_s-{1\over y} \left({\mu_t\over y}\right)_t
+{2\over (1-2\mu)} \left[(\mu_s+\mu)^2+(\mu_s+\mu)\mu
-{\mu_t^2\over y^2} \right]$$
$$\pm 2 \epsilon
\left[(\mu_s+\mu)^2 -{\mu_t^2\over y^2}\right]^{1/2}=
0,\eqno(3.1b)$$
$$\phi_s^2+(\phi_t/y)^2={1\over 2\pi}{\mu_s+\mu\over 1-2\mu},\eqno(3.1c)$$
$$2\phi_s(\phi_t/y)={1\over 2\pi}{\mu_t/y \over 1-2\mu}.\eqno(3.1d)$$
where $\epsilon=sgn(1-2\mu).$}

\vskip 0.3cm \noindent {\it Proof.} Let
$$X=\phi_s^2+(\phi_t/y)^2={1\over 2\pi}{\mu_s+\mu\over 1-2\mu},$$
$$Y=2\phi_s(\phi_t/y)={1\over 2\pi}{\mu_t/y \over 1-2\mu}.$$
It can be seen that either of the  conditions $\phi_s= \pm
\frac{\phi_t}{y}$ is equivalent to $\sqrt{X^2-Y^2}=0$. In the
following we  assume that $X^2-Y^2\ne 0$. Note that $\mu_t=0$
hence $Y=0$ can be achieved either by taking  $\phi_s=0$ or
$\phi_t=0$. These time independent solutions will be discussed in
Section 4. Assuming for example $\phi_s\ne 0$, we obtain
$$\phi_t/y={Y\over 2\phi_s}.$$
Substituting this in $X$ we have
$$\phi_s^4-X\phi_s^2+{Y^2\over 4}=0.$$
From this we can solve
$$\phi_s^2={1\over 2}[X\pm \sqrt{X^2-Y^2}].$$
Writing $$\Delta=X\pm \sqrt{X^2-Y^2}$$ we have
$$\phi_s=\pm{1\over \sqrt{2}} \Delta^{1/2},\quad\quad
  \phi_t/y=\pm{1\over \sqrt{2}}Y\Delta^{-1/2}.$$
  As noted at the end of Section 2, the signatures of $\phi_s$ and
$\phi_t$  are irrelevant, hence we may take the positive
signature.

 After substituting these and multiplying with $\Delta^{1/2}$,
the differential equation for $\phi$ reduces  to
$${1\over y}
\left[Y_t-{1\over 2} Y{\Delta_t\over \Delta}\right]= {1\over
1-2\mu}\Delta+{1\over 2}\Delta_s.$$ Substituting $\Delta$ and
taking derivatives one can see that the coefficients of $X_s$ and
$Y_t$ are proportional, in fact we have
$${1\over 2}[1\pm X(X^2-Y^2)^{-1/2}][Y_t/y-X_s]={1\over 1-2\mu}
                        (X\pm \sqrt{X^2-Y^2})
            \mp {1\over 2}
{ Y(Y_s-X_t/y) \over    (X^2-Y^2)^{1/2}}.$$ It can be seen that
$$Y_s-X_t/y=-{4\mu\over 1-2\mu}Y.$$
Hence it follows that
$$X_s-Y_t/y+{4\mu\over 1-2\mu}X\pm 2\sqrt{X^2-Y^2}=0$$
which is (3.1b). \hfill$\bullet$

For the rest of the paper we assume that $\mu_s+\mu>0$, hence we
take
 $\epsilon=1$ in the equation (3.1b).

\section{Static metrics: Exact solution for the positive branch}

In this section we shall study solutions for which the metric is
independent of $t$, hence static.  This will allow nevertheless a
time dependency in the scalar field. Putting $\mu_t=0$,
 the system of equations (3.1a-d)
reduces to
$${y_s\over y}=-2+{1\over 1-2\mu},\eqno(4.1a)$$
$$\mu_{ss}+\mu_s+{2\over 1-2\mu} (\mu_s+\mu)(\mu_s+2\mu)\pm
               2 (\mu_s+\mu)=0,\eqno(4.1b)$$
$$\phi_s^2+(\phi_t/y)^2={1\over 2\pi}{\mu_s+\mu\over 1-2\mu},\eqno(4.1c)$$
$$2\phi_s(\phi_t/y)={1\over 2\pi}{\mu_t/y \over 1-2\mu}.\eqno(4.1d)$$

We shall first give special solutions with  $\mu_s=k\mu$, then
obtain the exact solution for the positive branch and finally
study the solutions for the negative branch in Section 5.

\vskip 0.3cm \noindent {\bf  Special solutions with $\mu_s=k\mu$.}
\vskip 0.3cm

For the positive and negative branches we obtain respectively,

\hskip 1cm  {\it positive  branch:}$ \quad (k+1)(k+2)\mu=0,$

\hskip 1 cm {\it negative  branch:}$ \quad
(k+1)\left[8\mu^2+(k-2)\mu\right]=0.$

\noindent Exact solutions for these special cases are given below
with equations (4.2-5), where  $y_o$ and $\mu_o$ are integration
constants.

\vskip 0.2cm \noindent {\bf 1. $\mu=0$} is a special solution for
both positive and negative branches for which $\phi_s= \phi_t= 0$
and $y=y_oe^{-s}$. The line element is the flat metric
$$ds^2=-y_o^2\ dt^2 + \ dr^2 + r^2\ d\theta^2 + r^2
sin^2\theta \ d\varphi^2.\eqno(4.2)$$ \vskip 0.2cm \noindent {\bf
 2. $\mu=1/4$}
  is a solution for the negative  branch corresponding
to $\phi_s= 0$ and $\phi_t/y=1/2\sqrt{\pi} $. For this case $y$ is
a constant. Note that such a solution intersects the $m=r/2$ line
only at the origin. Thus its perturbations may lead to what is
observed as black holes with infinitesimal mass. The metric
$(2.1)$ becomes
$$ds^2=-2y_o^2  r^2\ dt^2 +2 \ dr^2 + r^2\ d\theta^2 + r^2
sin^2\theta \ d\varphi^2.\eqno(4.3)$$
 The only nonzero component
of the Ricci tensor is $R_{tt}=2y_o^2$ and the Ricci scalar is
$R=-\frac{1}{r^2}$. \vskip 0.2cm \noindent {\bf 3. $k=-1$}, or
$\mu_s+\mu=0$ corresponding to constant $m$ is a solution for
positive and negative branches. Here $\mu=\mu_oe^{-s}$ and
$y={y_o\over \mu_o} \mu|1-2\mu|$. This is the Schwarzschild
solution, it is the only possibility for intersecting the
$\mu=1/2$ line. For increasing $s$ the solution curves tend to the
origin. For negative $\mu$ they escape to infinity in the reverse
direction.  When $\mu_o$ is positive, it has  the standard
interpretation as the black hole mass. The metric (2.1) becomes
$$ds^2=-y_o^2\left[1-\frac{2\mu_o}{r}\right]\ dt^2 +
\left[1-\frac{2\mu_o}{r}\right]^{-1} \ dr^2 + r^2\ d\theta^2 + r^2
sin^2\theta \ d\varphi^2.\eqno(4.4)$$ \vskip 0.2cm \noindent {\bf
4. $k=-2$} or  $\mu_s+2\mu=0$ is a solution for the positive
branch. Then $\mu=\mu_oe^{-2s} $ and $y=y_o\mid{\mu\over
\mu_o}\mid ^{1/2}\ \mid 1-2\mu\mid^{1/2}$. For increasing $s$ the
curves tend to the origin. The part of the curve for positive
$\mu$ is not allowed. The part for negative $\mu$ escapes to
infinity in the reverse direction. The metric (2.1) becomes
$$ds^2=-y_o^2 dt^2 +
\left[1-\frac{2\mu_o}{r^2}\right]^{-1} \ dr^2 + r^2\ d\theta^2
 + r^2
sin^2\theta \ d\varphi^2.\eqno(4.5)$$ The only nonzero component
of the Ricci tensor is $R_{rr}=\frac{4\mu_o}{r^2(2\mu_o-r^2)}$ and
Ricci scalar is $R=-\frac{4\mu_o}{r^4}$.

\vskip 0.3cm \noindent {\bf Exact solution for the positive
branch.} \vskip 0.3cm For the positive branch $\phi_t= 0$ and we
write $\psi=\phi_s$.  The field equations  reduce to
$${y_s\over y}=-2+{1\over 1-2\mu}={4\mu-1\over 1-2\mu},\eqno(4.6a)$$
$$ \mu_s=-\mu+2\pi (1-2\mu)\psi^2 ,\eqno(4.6b)$$
$${\psi_s\over \psi}=-{1\over 1-2\mu}.\eqno(4.6c)$$
We will obtain an analytic solution for this system.

\vskip 0.3cm \noindent  {\bf \textit{Proposition 4.1.}} For
$\phi_t= 0$ the solution of the system (4.6a-c) is
$$\mu=\frac{1}{2}\left[\frac{\psi^2-\frac{c}{4\pi}\psi}
{\psi^2-\frac{c}{4\pi}\psi-\frac{1}{4\pi}}\right],\eqno(4.7)
$$
$${y\over y_o}= |\psi|\   |\psi-p|^{B_1}\ |\psi+q|^{B_2},\eqno(4.8)$$
$${r\over r_o}= |\psi|^{-1}\   |\psi-p|^{C_1}\ |\psi+q|^{C_2}\eqno(4.9)$$
where $\psi= \phi_s$, $r=e^s$, $y_o$ and $r_o$ are  constants,
$$p={1\over 8\pi}
 \left(c+ {\sqrt{c^2+16\pi}}\right)
={c\over 8 \pi}\left(1+{1\over \delta}\right),\eqno(4.10a)$$
$$q={1\over 8\pi} \left(-c+ {\sqrt{c^2+16\pi}}\right)
={c\over 8 \pi}\left(-1+{1\over \delta}\right) ,\eqno(4.10b)$$
$$B_1=-(1-\delta), \quad B_2=-(1+\delta),
\quad C_1={1\over 2} (1-\delta),\quad C_2={1\over
2}(1+\delta),\eqno(4.10c)$$ where $$\delta={c\over \sqrt{c^2
+16\pi}} .\eqno(4.11)$$ \vskip 0.3cm \noindent {\textit {Proof.}}
Let $u=(1-2\mu)\psi$. Taking the derivative of $u$ with respect to
$\psi$ and using (4.6b,c), we get
$$\frac{du}{d\psi}=u^2\left(4\pi+\frac{1}{\psi^2}\right).$$
This is a separable equation which  can be solved as
$$u=\frac{\psi}{1+c\psi-4\pi\psi^2}$$
where $c$ is an integration constant. From the definition of $u$
we obtain $\mu$ as a function of $\psi$ as in (4.7). Note that
there are vertical asymptotes at $\psi=p$ and $\psi=-q$, with $p$
and $q$ as in (4.10a,b). In the region $-q<\psi<p$, $\mu$ is less
than $1/2$ and has a unique maximum at $\psi_M=c/8\pi$ where
$$\mu_{M}=\mu(\psi_M)=\mu(c/8\pi)=\frac{1}{2}\frac{c^2}{c^2+16\pi}
={1\over 2}\delta^2.\eqno(4.12)$$ For $|c|>4\sqrt{\pi}$, $\mu$
crosses the level $\mu=1/2$ twice at $\psi^\pm$ given by
$$\psi^\pm={1\over 8\pi} \left(c\pm {\sqrt{c^2-16\pi}}\right).\eqno(4.13)$$
A typical solution curve is given in Figure 2.   Note that
$$\frac{d\mu}{d\psi}\mid_{\psi=0}=\frac{1}{2}c,$$
i.e., the
 parameter $c$ is the slope of the curve $\mu(\psi)$ at
$\psi=0$ or as $s\to \infty$. As we may assume $\psi>0$, we shall
restrict $\psi$ in the range $0<\psi<p$. For $c$  negative and
zero,  $\mu$ is negative, ranging from $-\infty $ to $0$ as $\psi$
changes from $p$ to $0$.  For $0<c<4\sqrt{\pi}$ it has a maximum
at $\psi_M $ but lies below the $\mu=1/4$ line.  As $c$ gets
large,  the solution curve is nearly a broken line following
closely the borders of the region bounded by $\mu=1/2$, $\psi=p$
and $\psi=-q$. For the critical value of $c=4\sqrt{\pi}$,
$\mu_{M}=1/4$ at $\psi_M=1/2\sqrt{\pi}$ and
$p=\frac{1}{2\sqrt{\pi}}(-1+\sqrt{2}).$ Typical solution curves
are presented in Figure 3a.

To solve for $y$ we write
$$\frac{dy}{y}=\frac{d\psi}{\psi}(1-4\mu)=\frac{d\psi}{\psi}
\frac{[-\psi^2+\frac{c}{4\pi}\psi-\frac{1}{4\pi}]}
{[\psi^2-\frac{c}{4\pi}\psi-\frac{1}{4\pi}]}=d\psi
\left[\frac{A}{\psi}+\frac{B_1}{\psi-p}+\frac{B_2}{\psi+q}\right]$$
where $A=1$ and $B_1$ and $B_2$ are as in (4.10c). By integrating
we obtain $y$ as in (4.8).
 It can be seen that
$dy/d\psi$ vanishes at $\psi^\pm$ for which $\mu(\psi^\pm)=1/4$.
Hence for $|c|\le 4\sqrt{\pi}$, $y$ is a monotone increasing
function of $\psi$ while  for  $|c|>4\sqrt{\pi}$, $y$ has two
critical points.
 As it can be seen
from Figure 4, for large $c$, the variation of $y$ with $\psi$ is
quite sharp.

Finally, substituting the expression of $\mu$ in (4.6c), we obtain
the equation
$${d\psi\over
4\pi\psi (\psi^2-\frac{c}{4\pi}\psi-\frac{1}{4\pi})} =d\psi
\left[\frac{A}{\psi}+\frac{C_1}{\psi-p}+\frac{C_2}{\psi+q}\right]=ds$$
where $A=-1$ and $C_1$ and $C_2$ are given by (4.10c). Integrating
this and exponentiating we obtain (4.9). It can be seen that
$r=e^s$ is a monotone decreasing function of $\psi$. Parametric
plots of $\mu$, $y$  and $r$ in terms of $\psi$, are shown in
Figure 4.\hfill $\bullet$ \vskip 0.3cm

From the remark above on the interpretation of $c$ and a
qualitative study of the solutions we see that the rate of change
of the mass distribution with respect to $\psi$ as $s\to \infty$
determines the solution completely. For large $c$, there is nearly
a black hole formation at  $s^*$ determined by
$\psi(s^*)=\psi_M=c/8\pi$, in the sense that, the solution curve
nearly touches the $\mu=1/2$ line but moves away from it to go
eventually to large negative values as $s\to -\infty$ or as $r\to
0$.

We were unable to find an exact solution for the negative branch.
For this case we shall study the equation (4.1b) as a dynamical
system.  To compare both branches and to analyze the efficiency of
numerical algorithms, we also studied the positive branch of
(4.1b) as a dynamical system in the $\mu$-$\mu_s$ plane. Writing
$\nu=\mu_s$, we can express the positive branch of (4.1b)
$$\left[
\begin{array}{c}
  \mu_s \\
  \nu_s \\
\end{array}
\right] = \left[
\begin{array}{c}
  \nu \\
  -3\nu-2\mu-{2\over 1-2\mu}
(\nu^2+3\mu\nu+2\mu^2) \\
\end{array}
\right].\eqno(4.14)$$ The only critical point is $(0,0)$ and the
linearized part at the origin is
$$\left[%
\begin{array}{c}
  \bar{\mu} \\
  \bar{\nu} \\
\end{array}%
\right]_s=\left[%
\begin{array}{cc}
  0 & 1 \\
  -2 & -3 \\
\end{array}%
\right]\left[%
\begin{array}{c}
  \bar{\mu} \\
  \bar{\nu} \\
\end{array}%
\right].\eqno(4.15)$$

\noindent The origin is a node and $\mu_s+\mu=0$ and
$\mu_s+2\mu=0$ are exact solutions. Typical solution curves shown
in Figure 5 are sketched using the exact solution. Since $\psi$ is
a monotone function of $s$, one can as well think of the phase
plane curves as parameterized by $\psi$ and make parametric plots
of $\mu$ and $\mu_s$. It can be seen that curves in the second
quadrant, lying between $\mu_s+\mu=0$ and $\mu_s+2\mu=0$
correspond to $c$ ranging from $-\infty$ to $0$, $\mu_s+2\mu=0$
being associated with $c=0$. The curves beyond $\mu_s+2\mu=0$
cross the $\mu=0$ line at a finite value of the parameter and
reach the origin as $s\to \infty$. The expressions of $\mu$ and
$\mu_s$ obtained from the exact solution implies that the origin
is a global attractor for all solutions starting in the region
defined by $\mu_s+\mu>0$ and $\mu<1/2$.

\section{ Static metrics:
Phase plane analysis for the  negative branch }

Writing $\nu=\mu_s$, we can express  second order equation (4.1b)
for $\mu$ as a dynamical system
$$\left[
\begin{array}{c}
  \mu_s \\
  \nu_s \\
\end{array}
\right] = \left[
\begin{array}{c}
  \nu \\
  \nu+2\mu-{2\over 1-2\mu}
(\nu^2+3\mu\nu+2\mu^2) \\
\end{array}
\right],\eqno(5.1)$$ where $\nu=\mu_s$. There are two critical
points, $(0,0)$ and $(1/4,0)$.    The linearized part at the
origin is
$$\left[%
\begin{array}{c}
  \bar{\mu} \\
  \bar{\nu} \\
\end{array}%
\right]_s=\left[%
\begin{array}{cc}
  0 & 1 \\
  2 & 1 \\
\end{array}%
\right]\left[%
\begin{array}{c}
  \bar{\mu} \\
  \bar{\nu} \\
\end{array}%
\right].\eqno(5.2)$$ To study the local behavior at $(0,0)$ we
note that the   eigenvalues of the system above  are $-1$ and $2$
hence the origin is a saddle point. The corresponding eigenvectors
are $(1,-1)$ and $(1,2)$, hence the solutions curves approaching
 the origin
along $\nu=-\mu$ in either direction are repelled and move away
along $\nu=2\mu$.

The linearized part at the point $(1/4,0)$ is
$$\left[%
\begin{array}{c}
  \bar{\mu} \\
  \bar{\nu} \\
\end{array}%
\right]_s=\left[%
\begin{array}{cc}
  0 & 1 \\
  -4 & -2 \\
\end{array}%
\right]\left[%
\begin{array}{c}
  \bar{\mu} \\
  \bar{\nu} \\
\end{array}%
\right].\eqno(5.3)$$
 The  eigenvalues of the system above are $-1\pm i\sqrt{3}$, hence the
point $(1/4,0)$ is a stable focus. The local behavior near the
points $(0,0)$ and $(1/4,0)$ will be  crucial in the proof of the
global behavior. We will now prove that $ (1/4,0)$ is a global
attractor  for solutions in the open half plane $\mu_s+\mu>0$,
$\mu<1/2$.

\vskip 0.3cm \noindent {\bf \textit{Proposition 5.1.}} \textit{Let
$(\mu,\nu)$ be a solution  of (5.1) and $D$ be the region bounded
by $\mu+\nu>0$ and $\mu<1/2$.  If $(\mu(0),\nu(0))$ belongs to
$D$, then the solution curve $(\mu(s),\nu(s))$  remains in $D$ for
all $s$ and $$\lim_{s\to\infty}(\mu(s),\nu(s))=(1/4,0).$$ } \vskip
0.3cm \noindent {\textit{Proof}}. Let $V$ be the velocity vector
of the solution curve. We can express $V$ as
$$V=\left[\nu,-{2\over (1-2\mu)}(\nu+2\mu)
(\nu+2\mu-{\textstyle{1\over 2}} \right].\eqno(5.4)$$ As $1-2\mu$
is positive on $D$, the direction of $V$ is determined by the
signatures of $\nu$, $\nu+2\mu$ and $\nu+2\mu-1/2$. This leads to
a subdivision of   $D$ into five regions as given  below. \vskip
0.3cm
\begin{tabular}{clcc }
Region    &              &    $\mu_s$&
$\nu_s$ \\
&&&\\
 $D_1$:& $\nu>0,\quad \mu+\nu>0,\quad  2\mu+\nu<0.$             &
$>0$& $<0$\\
$ D_2$:& $\nu>0,\quad 2\mu+\nu\geq0,\quad  1-4\mu-2\nu<0.$      &
$>0$&$>0$\\
$ D_3$:& $\nu>0,\quad 1-4\mu-2\nu\geq0,\quad \mu<1/2.$  &
$>0$&$<0$\\
$ D_4$:& $\nu\leq0,\quad 1-4\mu-2\nu>0,\quad  \mu<1/2.$ &
$\leq 0$&$<0$\\
$ D_5$:& $\nu\leq0,\quad \mu+\nu>0,\quad 1-4\mu-2\nu\leq0,\quad
\mu<\frac{1}{2}.$  &
$\leq0$&$>0$\\
&&&\\
\end{tabular}

\noindent In the region $D_1$, the velocity vector of the solution
curve lies in the fourth quadrant. By uniqueness it cannot
intersect the $\mu+\nu=0$ line. Furthermore as the origin is a
saddle point the solution curve necessarily intersects the
$\nu+2\mu=0$ line and enters in the region $D_2$.

The velocity vector of the solution curves in the region $D_2$ lie
in the second quadrant, hence each  solution curve passes to  the
region $D_3$ for increasing $s$.

In the region $D_3$, the velocity vector is in the fourth
quadrant. Here we will show that the boundaries $\nu+2\mu=1/2$ and
$\mu=1/2$ are never hit.

In order to show that solution curves do not hit the boundary
$\nu=-2\mu+1/2$, we prove that the slope of their velocity vector
is larger than $-2$.
 For this
we need the inequality
$$(\nu+2\mu)(\nu+2\mu-1/2)<\nu(1-2\mu)$$
to be satisfied. The equality in the relation above defines a
hyperbola, the half line $\nu+2\mu=1/2$, $\mu<1/4$  lies in the
region where the inequality is satisfied. Hence away from the
$\nu=0$ level, the velocity vector $V$ is surely pointing away
from the $\nu+2\mu=1/2$ line.  On the other hand as  $(1/4,0)$ is
a focus, the  solution curves cannot hit this point as $\nu$ gets
close to zero, hence necessarily, cross the vertical line
$\mu=1/4$.

To prove that solution curves do not hit the line $\mu=1/2$ at a
point $(\mu,\nu)$ with $\mu>1/4$, we need to show that the
velocity vector $V$ points below the line joining $(\mu,\nu)$ and
$(1/2,0)$, i.e., we need to prove that
$$\mid \nu_s/\mu_s\mid > \mid \nu/ ({1\over
2}-\mu)\mid.$$ Since $(1-2\mu)$ and $\mu_s=\nu$ are positive in
this region, we should  have
$$\mid\nu_s\mid\ (1-2\mu)-2\nu^2>0.$$
Substituting for $\nu_s$, we obtain
$$4\mu\nu+(2\nu+2\mu)(4\mu-1)>0.$$
For $\mu>1/4$ this inequality is surely satisfied and solution
curves reach the $\mu$ axis without hitting the $\mu=1/2$ line.

In the region $D_4$, the velocity vector belongs to the third
quadrant, hence the solution curves hit the $\nu+2\mu=1/2$ line
and pass to the region $D_5$. In the region $D_5$ the velocity
vector belongs to the second quadrant.  By uniqueness of
solutions, they cannot cross the $\mu+\nu=0$ line and the local
behavior  guarantees that they stay away from the origin and cross
the $\mu$ axis. Again by uniqueness the solution curves spiral
around the point $(1/4,0)$ which is a stable focus.
\hfill$\bullet$

\vskip 0.3cm The integral curves for the numerical solution of the
dynamical system 5.1 in the domain $\mu_s+\mu>0$ and $\mu<1/2$ are
given in Figure 7. Expect for the line $\mu_s+\mu=0$, all solution
curves in this region spiral around the point $(1/4,0)$ as $s\to
\infty$. As one moves to right, the solution curves get very close
to the line $\mu=1/2$ but decay quickly to the level $\mu=1/4$.
The variations of $\mu,$ $y$ and $\phi_t/y$ with respect to $s$
exhibit rapidly decaying oscillations as shown in Figure 8.

The period of the linearized part of the dynamical system (5.1) at
the point $(\frac{1}{4},0)$ is $\frac{2\pi}{\sqrt{3}}\simeq 3.62$.
For the nonlinear system we could observe only 3 periods of
oscillations. Starting with the initial conditions
$(\mu=0,\mu_s=\mu_s^0)$, $\mu_s^0$ ranging from $10^{-7}$ to
$10^3$, we computed consecutive crossings of the integral curves
across $\mu=1/4$ line at $\mu_s=a_i$, $i=1,...,4$ and we evaluated
the periods $T_i$, $i=1, 2, 3$ as the distance $s$ between two
consecutive crossings. Representative values for $a_i$'s and
$T_i$'s are listed below. \vskip 0.2cm
\begin{tabular}{ c c c c c}
  $\mu_s^0=10^{-7}:$& $a_1, a_2, a_3$ & 0.0923 & 0.0022 & 0.0001\\

                   &$T_1, T_2, T_3$  &3.5350   &3.6280   &3.6070\\
\\
  $\mu_s^0=1$      :& $a_1, a_2, a_3$&0.4698  &0.0089  &  0.0002\\

                   & $T_1, T_2, T_3$ & 3.3510 & 3.6180 &3.6380\\
\\
  $\mu_s^0=10^3$   :& $a_1, a_2, a_3$&567.4564  & 0.0837  & 0.0020\\

                   & $T_1, T_2, T_3$ & 4.7040 & 3.5430  &3.6280\\

\end{tabular}\\

\noindent As $\mu_s^0$\quad goes to infinity the first crossing
level $a_1$ goes to infinity i.e.,
$\lim_{\mu_s^0\rightarrow\infty} a_1=\infty$, while for $i>1$ they
tend to a limit that we denote $a_i^*$, with
$$\lim_{\mu_s^0\rightarrow\infty}
a_{i+1}=a_i^*=\lim_{\mu_s^0\rightarrow 0} a_i.$$ For $i=1,2,3$, we
have $a_1^*\cong 0.092,\quad a_2^*\cong 0.0022,\quad a_3^*\cong
0.0001.$

Although the first oscillation depends on the initial condition
the second oscillation start near $\mu_s=0.0022-0.0090$ and have a
quite stable period of $3.62$, close to the period of the
linearized system.

\vskip 0.2cm \noindent{\bf{Remark 5.1.}} Putting $\mu_{tt}=k\mu_s$
in the linearized part of the full system and writing
$\mu=\frac{1}{4}+u$ we have
$$u_{ss}+(2-k)u_s+4u=0.$$
For $k=2$, the solution of linearized and nonlinear system are
unrelated as predicted by Hartman-Grobman theorem \cite{HG}. For
arbitrary $0<k<2$ the linearized solution is of the form
$u=e^{-\alpha s}e^{\pm i\omega s}$ where $\alpha=1-\frac{k}{2}$
and $\omega^2+\alpha^2=4.$ As an interesting coincidence we
observed that the period corresponding  to $k=0.3704$ is
$T=\frac{2\pi}{1.8265}=3.44$.

\ack The authors would like to thank Dr. D. Grumiller for the
pointing out references \cite{F48}-\cite{Br02} and Professor B.K.
Harrison for illuminating discussions. This work is partially
supported by the Turkish National Council for Scientific and
Technological Research.

\newpage
\begin{figure}
\leavevmode
\rotatebox{0}{\scalebox{0.5}{\includegraphics{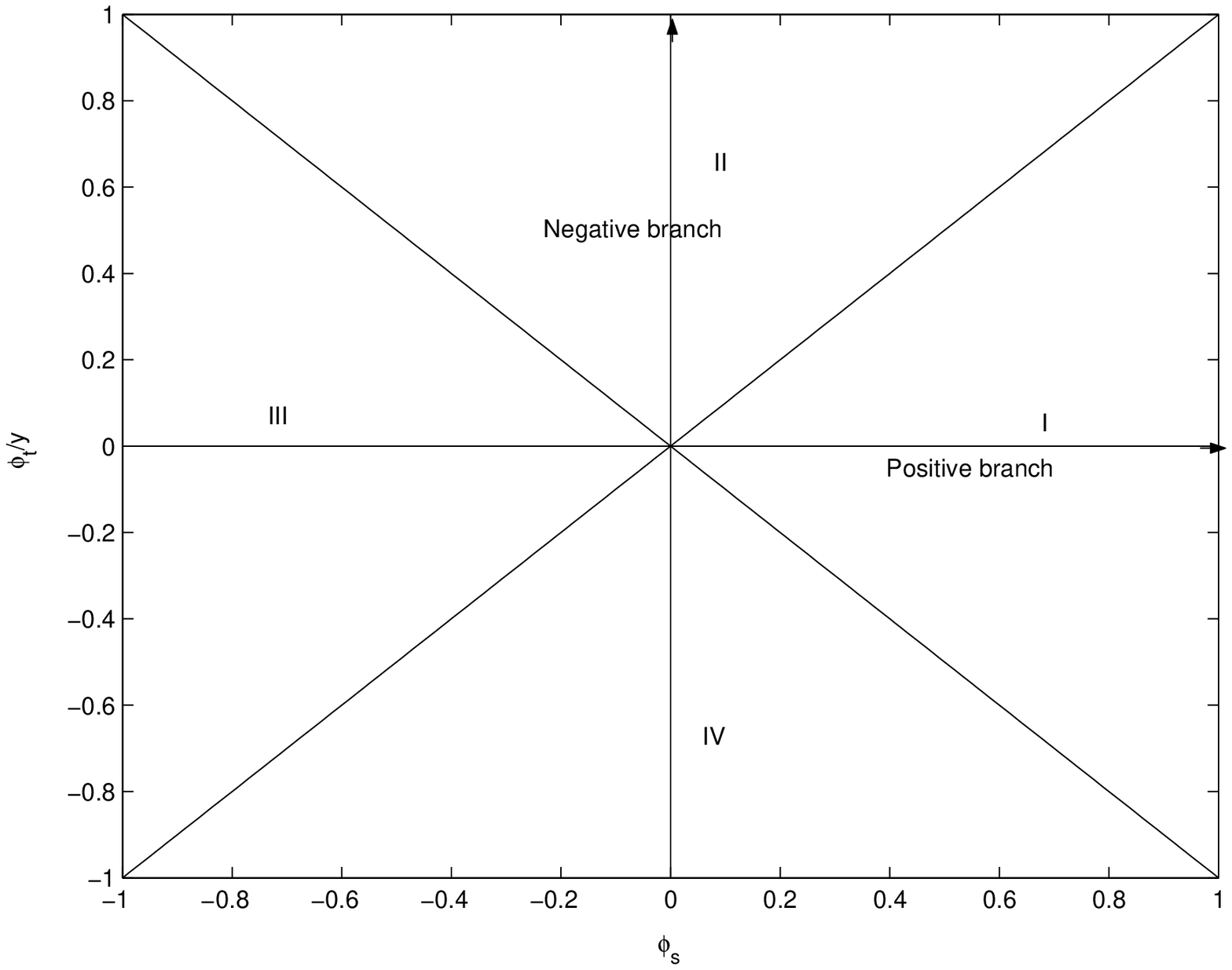}}}
\\
{{\bf Figure 1a.}
 The regions determined by the conditions
$\phi_s=\pm {\phi_t\over y}$.  Since $\phi_s\rightarrow-\phi_s$ is
a discrete symmetry,  it is sufficient to work with $\phi$ in
regions $I$ and $II$.}
\end{figure}

\begin{figure}
\leavevmode
\rotatebox{0}{\scalebox{0.5}{\includegraphics{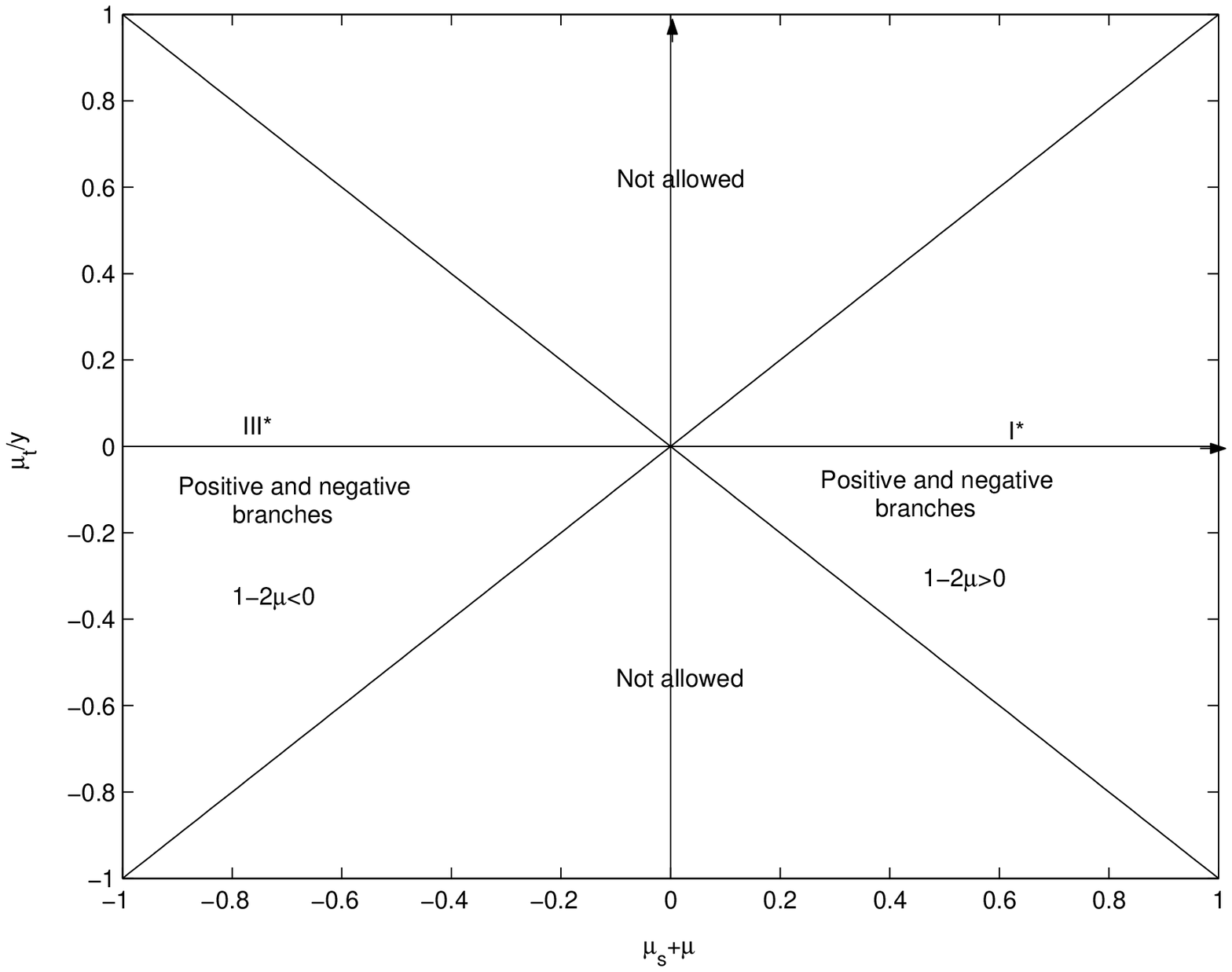}}}
\\
{{\bf Figure 1b.} The regions determined by the conditions
$\mu_s+\mu=\pm {\mu_t\over y}.$ Assuming $1-2\mu>0$, only region
$I^*$ is allowed.}
\end{figure}

\begin{figure}
\leavevmode
\rotatebox{0}{\scalebox{0.5}{\includegraphics{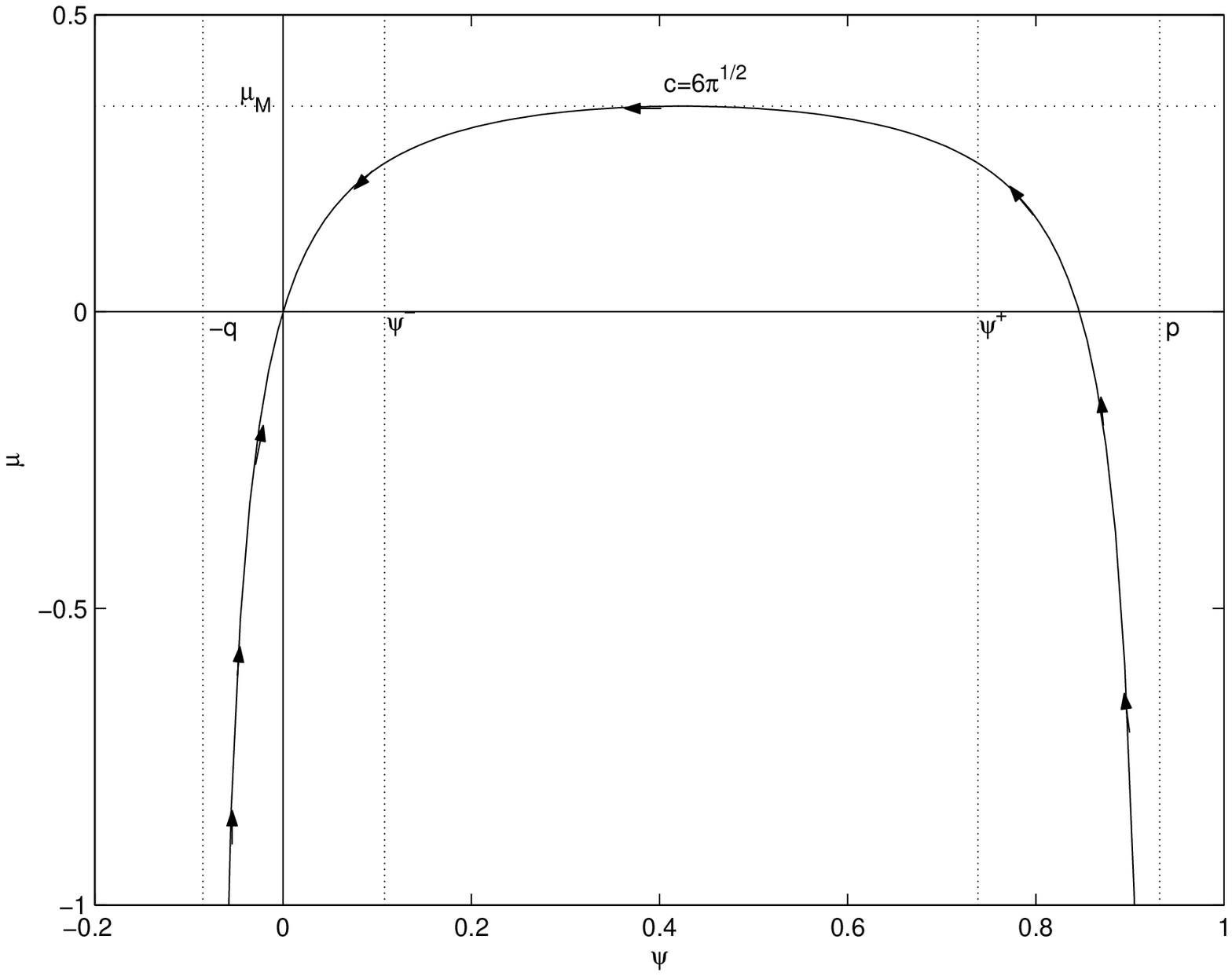}}}
\\
{{\bf Figure 2.} General behavior of $\mu$ for a parameter value
$c=6\sqrt{\pi}$ in the domain $-q<\phi_s=\psi<p$.  The values $p$,
$q$,  $\psi^\pm$ and $\mu_M$ are defined by Eqns. (4.10-13). }
\end{figure}

\begin{figure}
\leavevmode
\rotatebox{0}{\scalebox{0.5}{\includegraphics{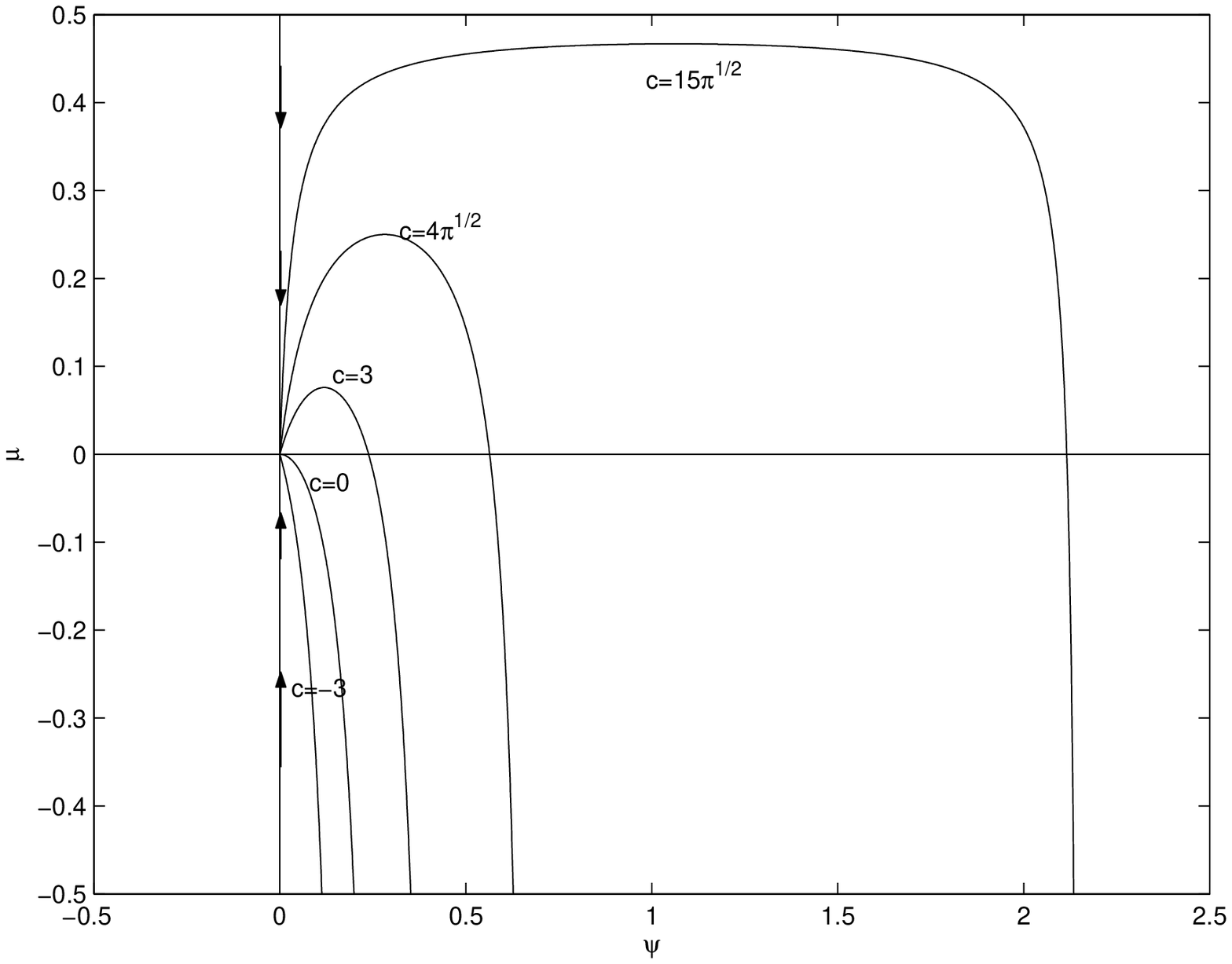}}}
\\
{{\bf Figure 3a.} Variation of $\mu$ with $\psi=\phi_s$  for
different values of $c$, for  $0<\psi<p$.}
\end{figure}

\begin{figure}
\leavevmode
\rotatebox{0}{\scalebox{0.5}{\includegraphics{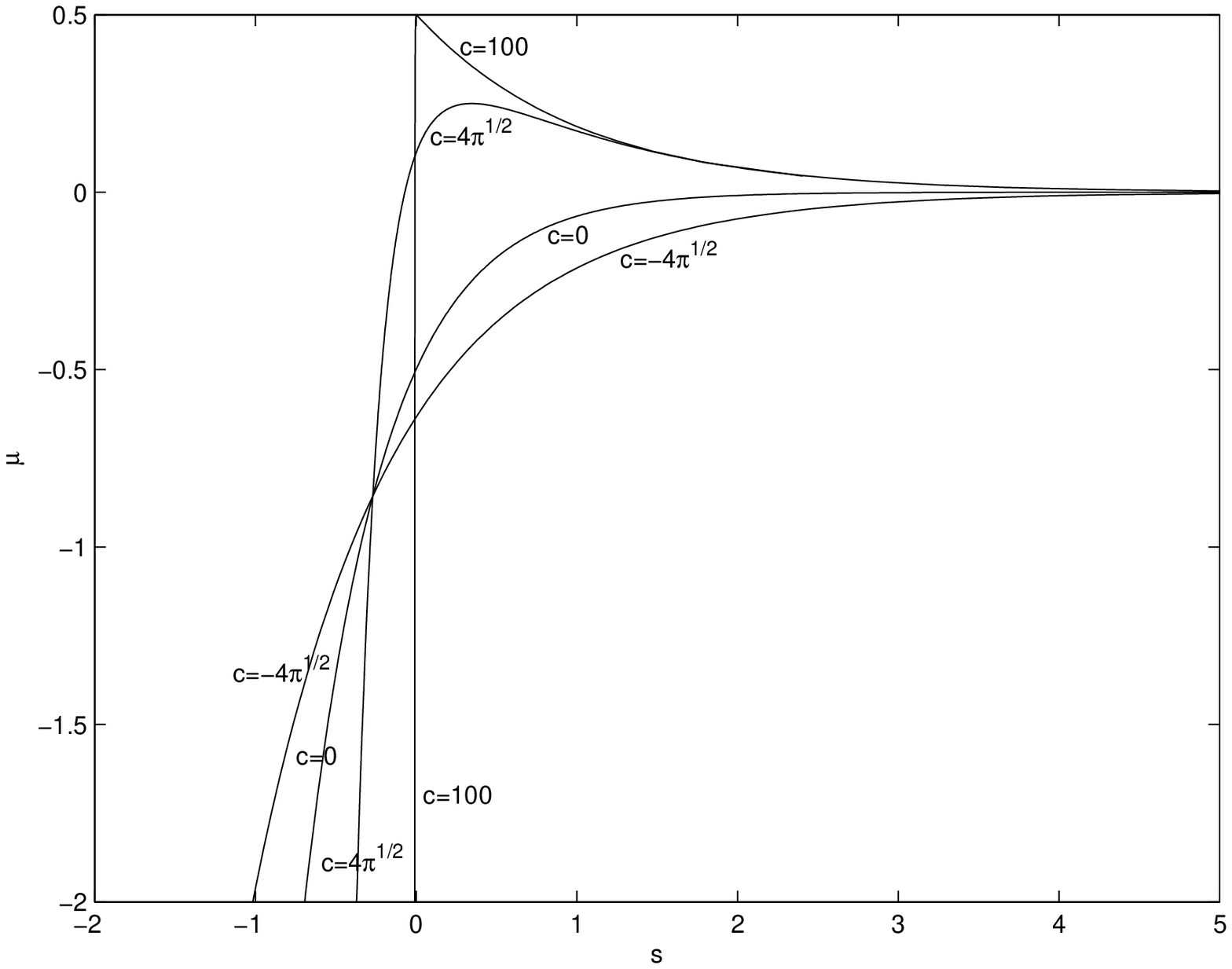}}}
\\
{{\bf Figure 3b.} Variation of $\mu$ with $s$ for different values
of $c$ for $-\infty<s<\infty$.  For $c\le 0$, $\mu$ is monotone
while for $c>0$ it has a unique maximum $\mu_M$ at a finite $s$.
For $c$ large, $\mu(s)$ makes a sharp peak at this critical point,
as opposed to the nearly flat variation of $\mu$ with respect to
$\phi_s$ as seen in Figure 3a. }
\end{figure}

\begin{figure}
\leavevmode
\rotatebox{0}{\scalebox{0.5}{\includegraphics{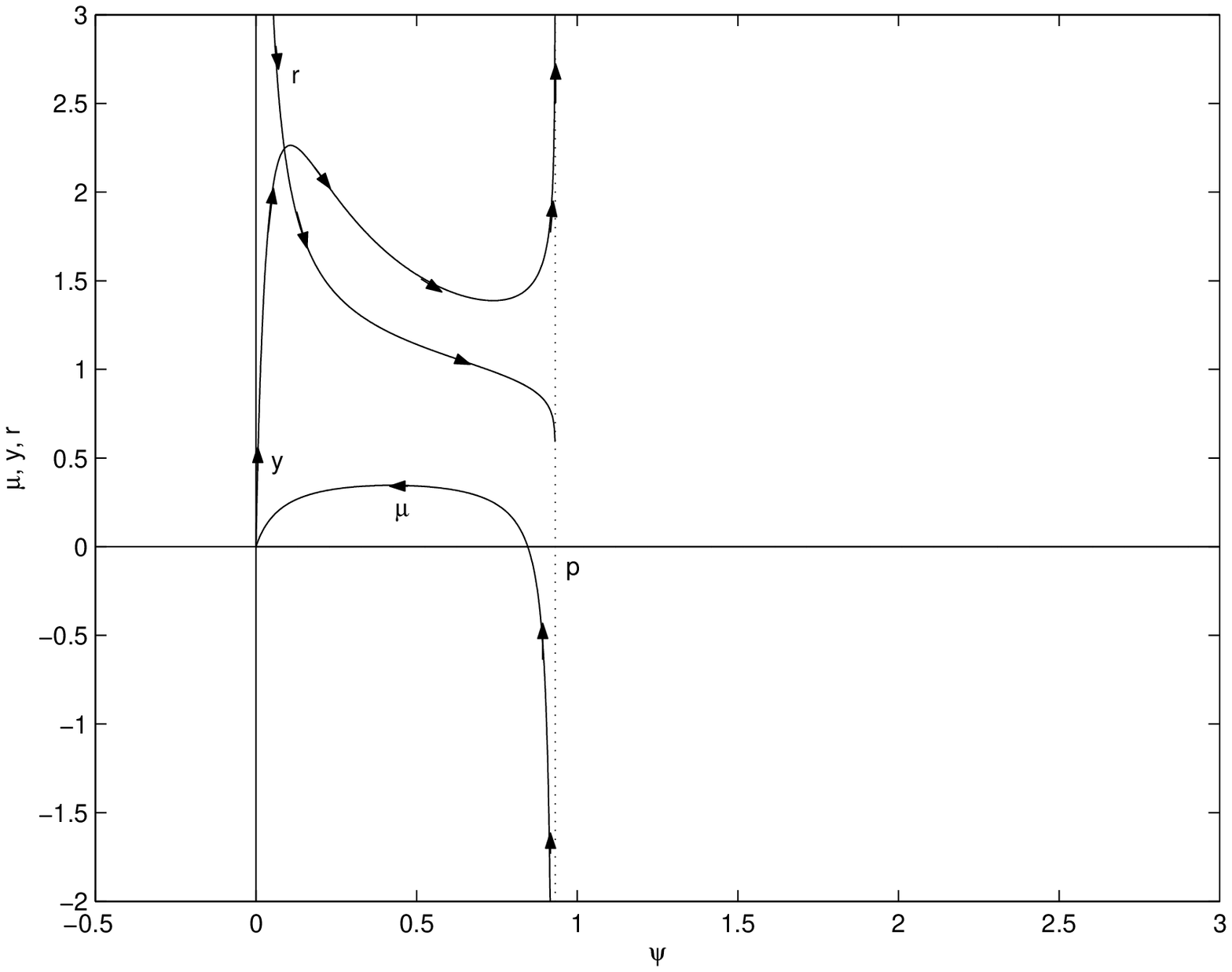}}}
\\
{{\bf Figure 4.} The variations of $\mu$, $y$ and $r$ with respect
to $\psi=\phi_s$ for  $c=6\sqrt{\pi}$ and $0<\psi<p$.  Recall that
for $c\le 0$,  $\mu$ is monotone but has a unique maximum for
$c>0$. On the other hand, $y$ is monotone for $c\le 4\sqrt{\pi}$
but has two critical points for $c>4\sqrt{\pi}$ while $\psi$ is
monotone decreasing for all  $c$.
 We have chosen
the initial value of $y$ as $y_o=1$. For increasing $s$, $\mu$
rises to its maximum $\mu_M$ and decays to zero, while $y$ starts
from its initial value reaches its maximum, decays nearly to zero
especially for large $c$ and finally rises to $\infty$ very
sharply.  }
\end{figure}

\begin{figure}
\leavevmode
\rotatebox{0}{\scalebox{0.5}{\includegraphics{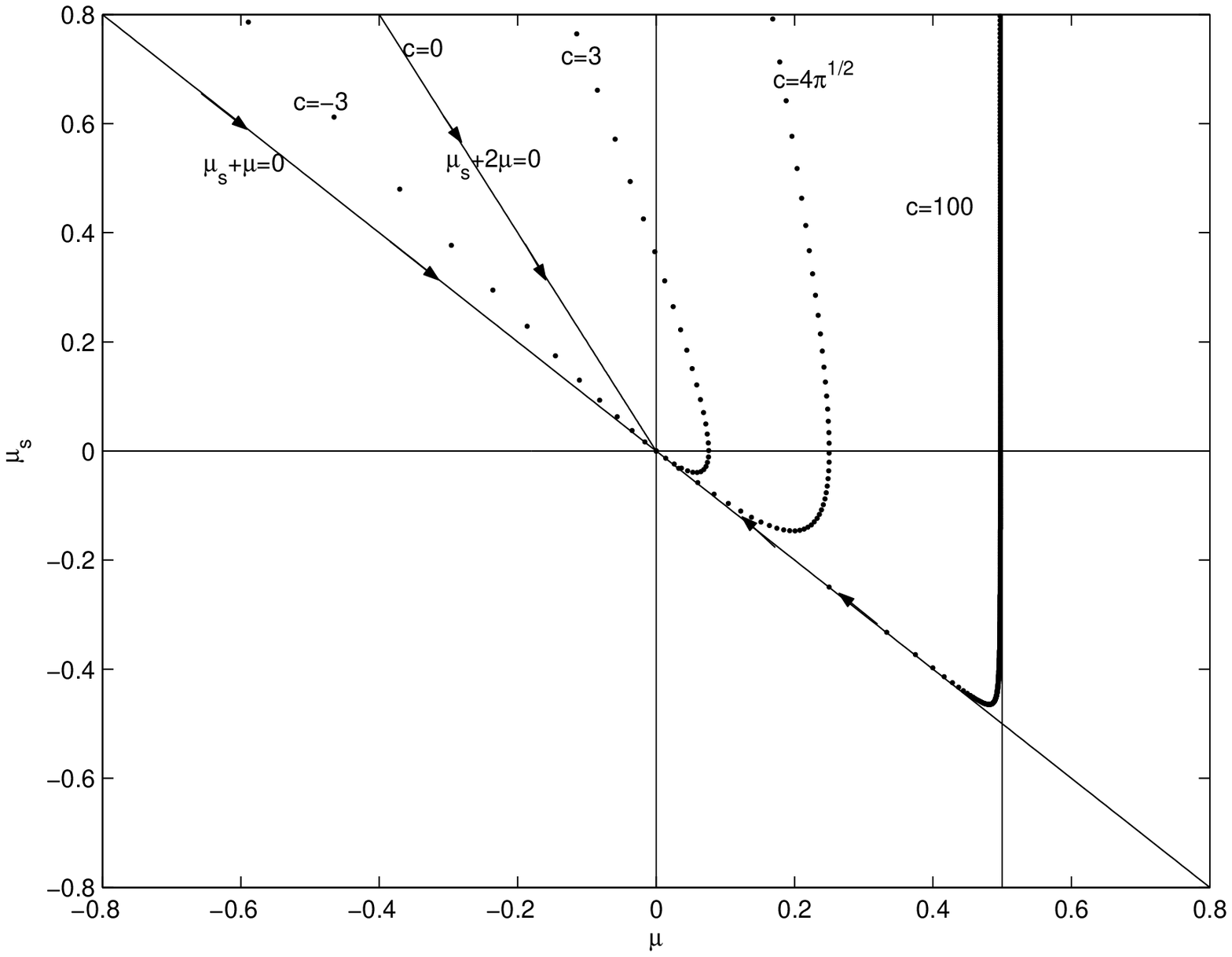}}}
\\
{{\bf Figure 5.} The integral curves of the dynamical system
(4.14) for the positive branch  in the domain $\mu_s+\mu>0$ and
$\mu<1/2$.  The lines $\mu_s+\mu=0$ and $\mu_s+2\mu=0$ are exact
solutions.  The half line $\mu_s+\mu=0$, $\mu<0$ corresponds to
the limit $c\to-\infty$. For the solution curves in between this
half line and $\mu_s+2\mu=0$, $c$ ranges from $-\infty $ to zero.
For these solutions $\mu$ is negative, and monotone increasing and
reach the origin as $s\to\infty$. For $c$ positive, all solution
curves reach their maximum value $\mu_M<1/2$ and arrive at the
origin as $s\to \infty $. Note the sharp changes near the
$\mu=1/2$ line for large $c$.}
\end{figure}

\begin{figure}
\leavevmode
\rotatebox{0}{\scalebox{0.5}{\includegraphics{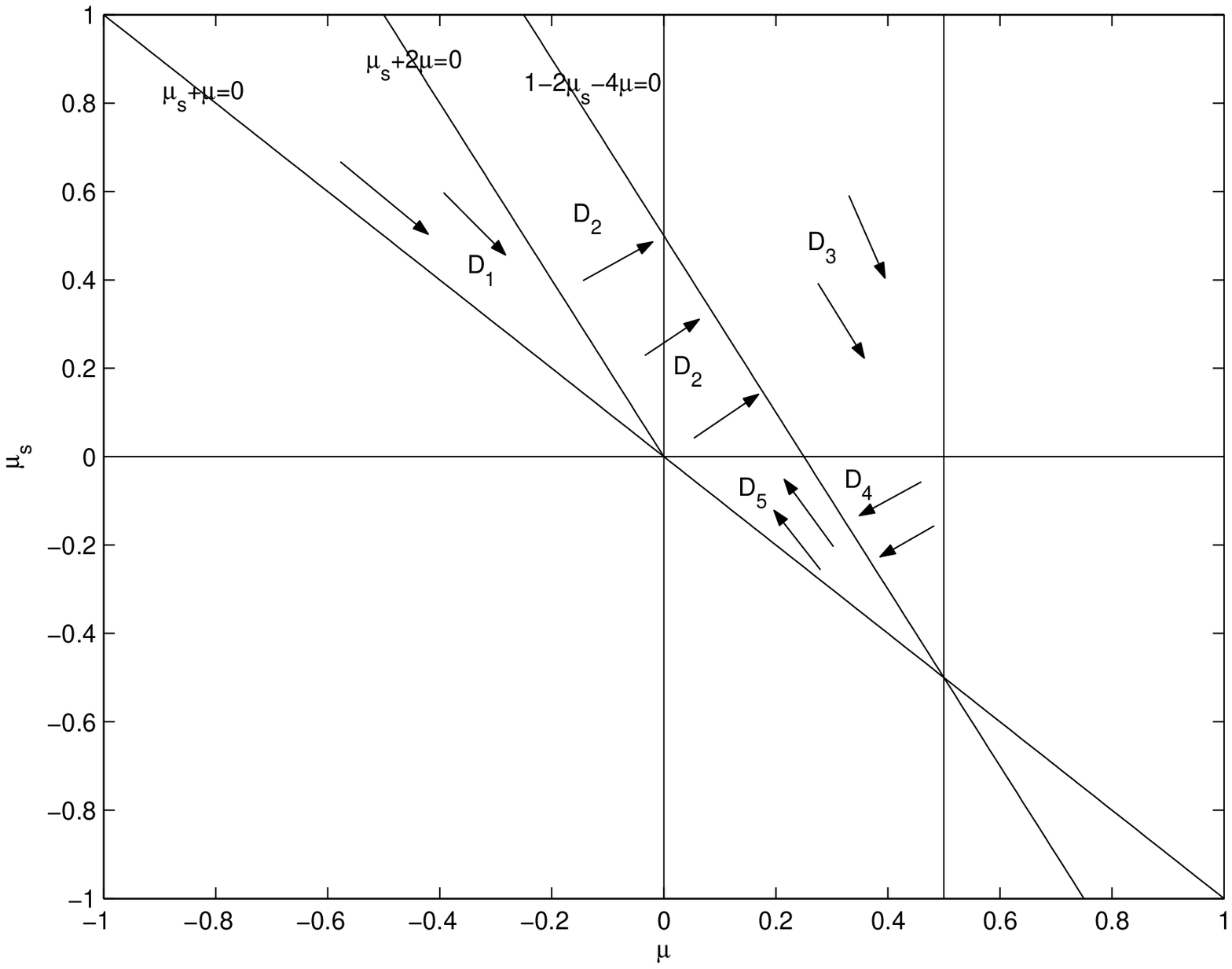}}}
\\
{{\bf Figure 6.} The direction fields for the dynamical system
(5.1) for the negative branch in the regions $D_i$, $i=1,\dots 5$
for $\mu_s+\mu>0$ and $\mu<1/2$. }
\end{figure}

\begin{figure}
\leavevmode
\rotatebox{0}{\scalebox{0.5}{\includegraphics{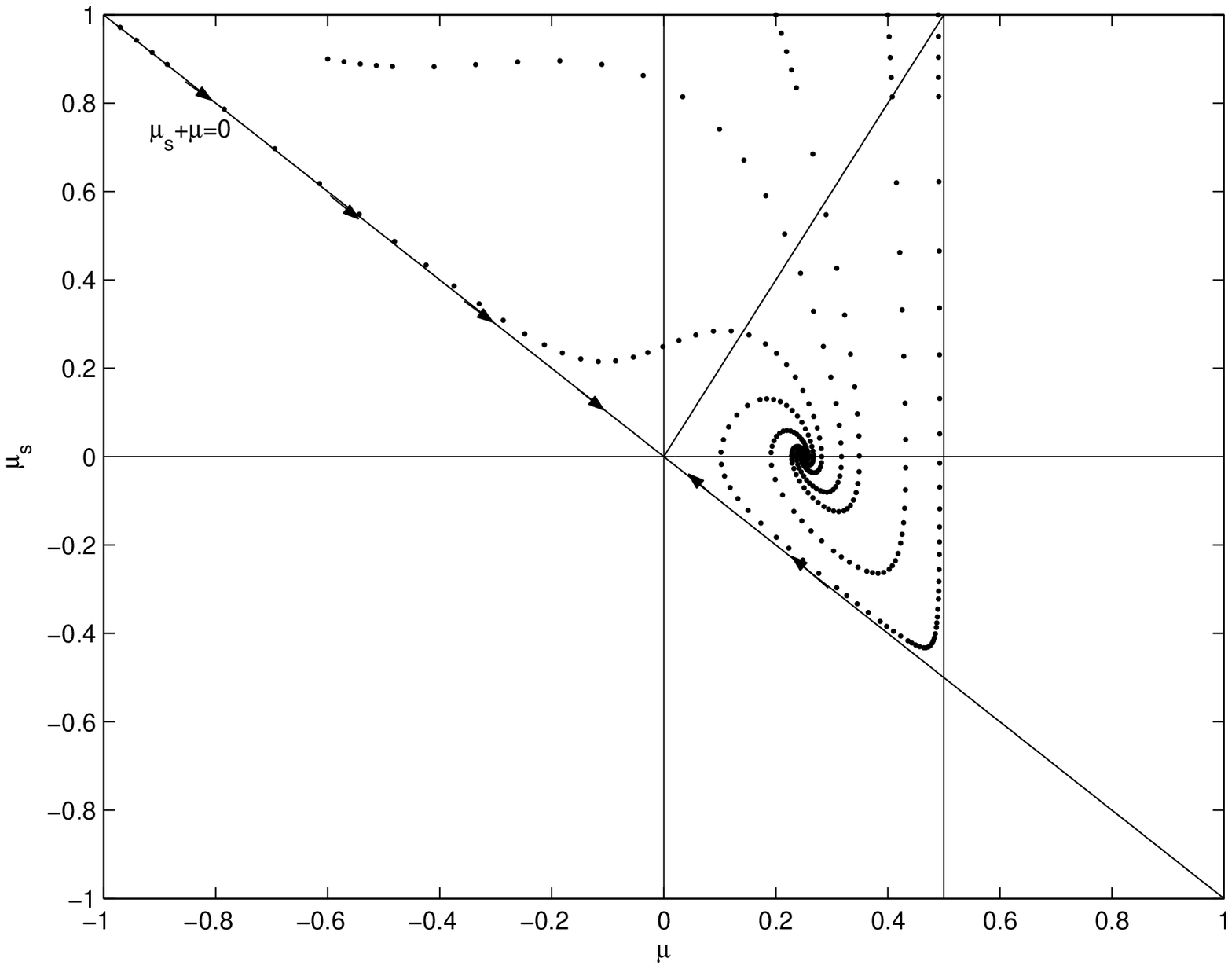}}}
\\
{{\bf Figure 7.} The integral curves of the dynamical system (5.1)
for the negative branch  in the domain $\mu_s+\mu>0$ and
$\mu<1/2$. The line $\mu_s+\mu=0$ is an  exact solution. All
solution curves in this region spiral around the point $(1/4,0)$
as $s\to \infty$. The curves sketched here are determined by their
initial values
 $(\mu(0),\mu_s(0))$, which are respectively   $(-1,1.001)$,
$(-0.6,0.9)$, $(0.2,1)$, $(0.4,1)$, $( 0.49,1)$ respectively from
right to left. Note that as one moves to right,
 the solution curves get very close to the line $\mu=1/2$ but
 decay quickly to the level $\mu=1/4$ making small amplitude
 oscillations around this level.}
\end{figure}

\begin{figure}
\leavevmode
\rotatebox{0}{\scalebox{0.5}{\includegraphics{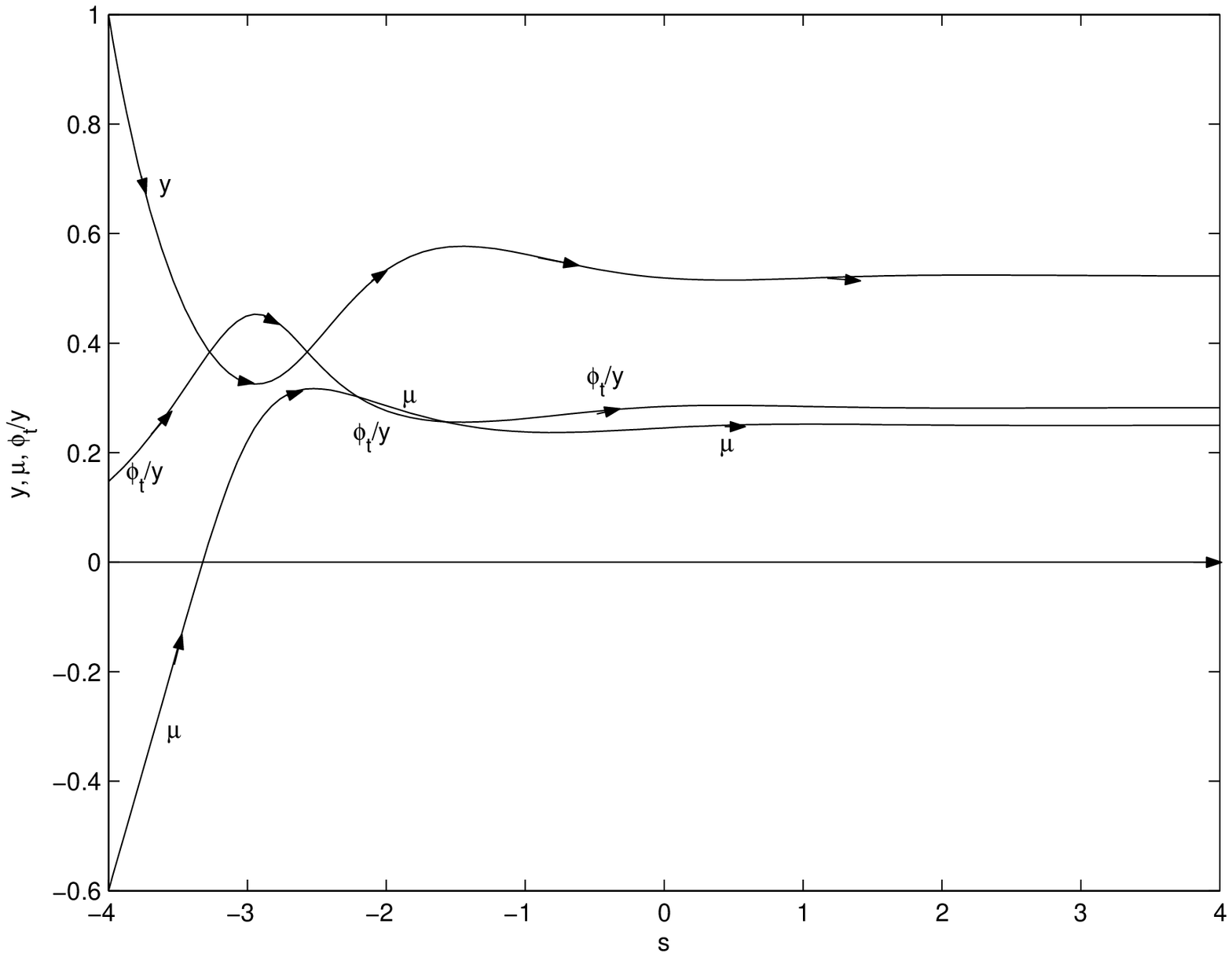}}}
\\
{{\bf Figure 8.} The variations of $y$, $\mu$ and $\phi_t/y$ with
respect to $s$ for  the dynamical system (5.1) and Eqn.(2.6a). The
curves sketched here are determined by their initial value
 $(y(0),\mu(0),\mu_s(0))$, which is $(1,-0.6,0.9)$. Note that the solution of $\phi_t/y$ are
 obtained from Eqn.(2.6c) for $\phi_s=0$.}
\end{figure}

\end{document}